\documentclass[twocolumn,showpacs,preprintnumbers,aps,pra,amsmath,amssymb,floatfix,superscriptaddress]{revtex4-1}

\usepackage{graphicx}% Include figure files
\usepackage{subfigure}
\usepackage{dcolumn}% Align table columns on decimal point
\usepackage{bm}% bold math
\usepackage{color}
\usepackage{natbib}
\usepackage{textcomp}
\usepackage{url}           % for on-line citations
\usepackage{appendix}
\usepackage{hyperref}
\usepackage{tabularx}
\hypersetup{colorlinks=true,urlcolor=blue,breaklinks=true,citecolor=blue,linkcolor=blue,pdfstartview=FitH,pdfpagemode=UseNone} 

\newcolumntype{C}{>{\centering\arraybackslash}X}

\newcommand{\ket}[1]{\left|{#1}\right\rangle}
\def\Rb{$^{87}$Rb }
\def\Na{$^{23}$Na }
\def\NaRb{$^{23}$Na$^{87}$Rb }

\begin{document}

\title{High-resolution Internal State Control of Ultracold $^{23}$Na$^{87}$Rb Molecules}

\author{Mingyang Guo}
\email{myguo@phy.cuhk.edu.hk}
\affiliation{Department of Physics, The Chinese University of Hong Kong, Shatin, Hong Kong, China}
\author{Xin Ye}
\affiliation{Department of Physics, The Chinese University of Hong Kong, Shatin, Hong Kong, China}
\author{Junyu He}
\affiliation{Department of Physics, The Chinese University of Hong Kong, Shatin, Hong Kong, China}
\author{Goulven Qu\'{e}m\'{e}ner}
\affiliation{Laboratoire Aim\'{e} Cotton, CNRS, Universit\'{e} Paris--Sud, ENS Paris--Saclay, Universit\'{e} Paris--Saclay, 91405 Orsay Cedex, France}
\author{Dajun Wang}
\email{djwang@cuhk.edu.hk}
\affiliation{Department of Physics, The Chinese University of Hong Kong, Shatin, Hong Kong, China}
\affiliation{Shenzhen Research Institute, The Chinese University of Hong Kong, Shenzhen, Guangdong, China} 
%\affiliation{The Chinese University of Hong Kong Shenzhen Research Institute, Shenzhen, China} 

\date{\today}

\begin{abstract}

We report the full control over the internal states of ultracold \NaRb molecules, including vibrational, rotational and hyperfine degrees of freedom. Starting from a sample of weakly bound Feshbach molecules, we realize the creation of molecules in single hyperfine levels of both the rovibrational ground and excited states with a high-efficiency and high-resolution stimulated Raman adiabatic passage. Starting from the rovibrational and hyperfine ground state, we demonstrate rotational and hyperfine control with one- and two-photon microwave spectroscopy. This achievement of fully controlling the molecular internal states paves the way to study state dependent molecular collisions and state controlled chemical reactions.

\end{abstract}

\maketitle

%%%%%%%%%%%%%%%%%%%%%%%%%%%%%%%%%%%%%

Ultracold polar molecules (UPMs) have long been predicted to have great potential applications in ultracold chemistry~\cite{krems2008cold}, quantum simulation of novel many-body physics~\cite{carr2009cold,trefzger2011ultracold,baranov2012condensed}, and quantum computation~\cite{demille2002quantum}. In recent years, with more and more ultracold molecular species of different chemical reactivity and quantum statistics being created~\cite{ni2008high,takekoshi2014ultracold,molony2014creation,park2015ultracold,guo2016creation}, some of these predictions are becoming experimental reality. In the pioneer $^{40}$K$^{87}$Rb experiment of JILA, ultracold chemical reaction was observed and controlled by dipolar interaction and dimensionality~\cite{ospelkaus2010quantum,ni2010dipolar,miranda2011controlling}, a lattice gas of UPMs with filling factor above the percolation threshold was created~\cite{moses2015creation,covey2016doublon}, and many-body dipolar spin-exchange in optical lattices was observed~\cite{yan2013observation,hazzard2014many}.  

A prerequisite for all of these achievements is a sample of UPMs with well controlled internal states satisfying specific application requirements. Because of the permanent dipole moments of UPMs, a very convenient method for rotational and hyperfine state control is microwave spectroscopy, which has been successfully implemented in several species~\cite{ospelkaus2010controlling,will2016coherent,gregory2016controlling,park2017second}. Coupling rotational levels coherently with microwave is an important way of inducing dipolar interactions between molecules for realizing novel lattice quantum magnetism models~\cite{barnett2006quantum,gorshkov2011tunable,yan2013observation,hazzard2014many} and topological phases~\cite{cooper2009stable}. Microwave dressing can also be combined with static electric fields to engineer long-range repulsive barriers for suppressing inelastic collisions~\cite{gorshkov2008suppression}. In addition, at some particular electric fields, UPMs created at the $J=1$ rotational level  (with $J$ the rotational quantum number) with the help of microwave spectroscopy are predicted to have enough favorable elastic to inelastic collision ratios for evaporative cooling~\cite{avdeenkov2006suppression,wang2015tuning,quemener2016shielding,gonzalez2017adimensional}.   

In this paper, we adopt the microwave spectroscopy to control the rotational and hyperfine states of the ultracold ground-state \NaRb molecules, which have been created in our group~\cite{guo2016creation}. In addition, we also demonstrate hyperfine resolved control of the vibrational and rotational states directly via the two-photon Raman population transfer. Vibrational excitation has been established as an efficient technique to modify the rate and product state distribution of chemical reactions since the 1970s~\cite{odiorne1971molecular,polanyi1972concepts,zare1998laser}. Importantly, for the several non-reactive UPMs recently produced in the lowest vibrational state ($v = 0$, with $v$ the vibrational quantum number)~\cite{zuchowski2010reactions}, including $^{87}$Rb$^{133}$Cs~\cite{takekoshi2014ultracold,molony2014creation}, $^{23}$Na$^{40}$K~\cite{park2015ultracold} and $^{23}$Na$^{87}$Rb~\cite{guo2016creation}, the chemical reaction can be activated if the first excited vibrational state ($v=1$) is populated instead. This actually enabled us to investigate the ultracold molecular collisions with controlled chemical reaction~\cite{ye2017collisions}.

\begin{figure}[bpt]
	\centering
	\includegraphics[width=0.45\textwidth]{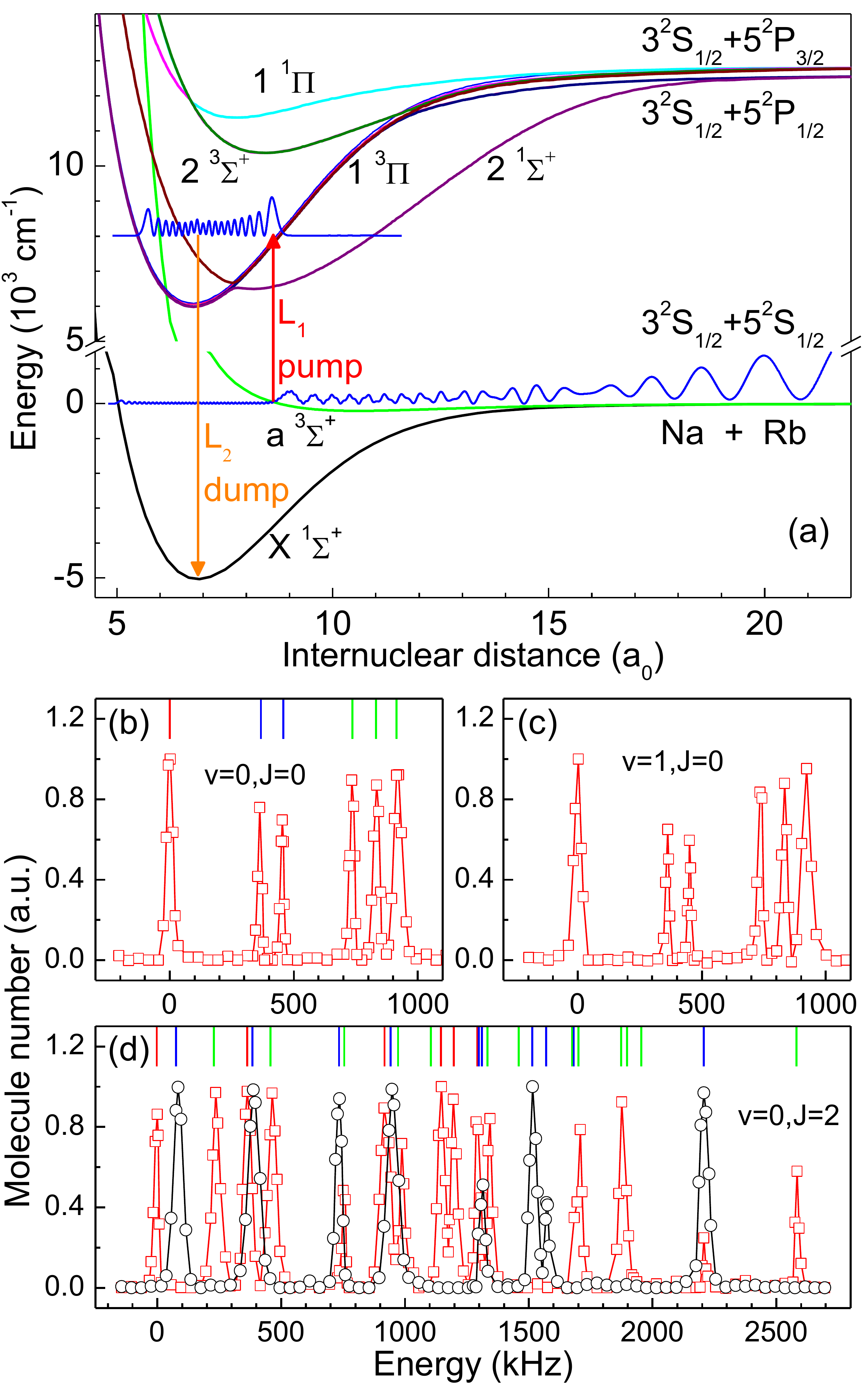}
	\caption{Hyperfine resolved internal state control with Raman lights. (a) shows the \NaRb potential energy curves and the two-photon Raman process for population transfer. (b), (c) and (d) are STIRAP spectra of the ($v=0, J=0$), ($v=1, J=0$) and ($v=0, J=2$) states of the $X\,^1\Sigma^+$ potential obtained by tuning the frequency of $L_2$. The energy zeros of (b), (c) and (d) correspond to binding energies of 149.20997(9)~THz, 146.03718(9)~THz and 149.19747(9)~THz relative to two free atoms both in $\ket{F=1,M_F=1}$ states at 335.2~G, with the uncertainties determined by the wavelength meter used for absolute frequency measurement. In (d), the spectrum in black circles is obtained with $L_2$ of $\pi$ polarization, and the one in red squares is obtained with $\sigma^\pm$ polarization. (b) and (c) are obtained with mixed $L_2$ polarizations. The color-coded vertical bars in (b) and (d) indicate the calculated positions of the hyperfine levels using the fitted coupling constants in Table~\ref{table1} with $M_F=3$ in red, $M_F=2$ in blue and $M_F=1$ in green.}
	\label{figure1}
\end{figure}

The experiment starts from a pure sample of weakly bound \NaRb Feshbach molecules obtained via magneto-association with a 347.6~G Feshbach resonance between $^{23}$Na and $^{87}$Rb atoms both in their $\ket{F = 1,M_F=1}$ hyperfine Zeeman states ~\cite{wang2013observation,wang2015formation}. Here, $F$ is the total atomic angular momentum, while $M_F$ is its projection along the quantization axis provided by the magnetic field, which is along the vertical direction and has a strength of $B = 335.2$~G after the magneto-association. The sample is trapped in a cigar-shaped optical potential formed by two crossed 1064.4~nm laser beams. A stimulated Raman adiabatic passage (STIRAP) is then applied to transfer the molecules to target rovibrational levels of the $X\,^1\Sigma^+$ potential. Figure~\ref{figure1}(a) shows the two-photon Raman process schematically, in which the pump light $L_1$ couples the Feshbach state and the intermediate state while the dump light $L_2$ couples the intermediate state and the final rovibrational state. $L_1$ and $L_2$ copropagate perpendicularly to the magnetic field. The two Raman lasers are locked to the same ultra-stable high-finesse optical cavity using the Pound-Drever-Hall technique~\cite{drever1983laser} with their linewidths narrowed down to 1~kHz.

%%%%%%%%%%%%%%%%%%%%%%%%%%

The intermediate state for the STIRAP, which is the same one used to produce the absolute ground-state \NaRb molecule, is a singlet/triplet mixed $2\,^1\Sigma^+-1\,^3\Pi$ level with about 5\% $2\,^1\Sigma^+$ character~\cite{guo2016creation,guo2017high}. Thanks to the favorable Franck-Condon overlap, this level has strong transition strengths to a series of vibrational levels near the bottom of the $X\,^1\Sigma^+$ potential [Fig.~\ref{figure1}(a)]. Indeed, by tuning the laser frequencies, we successfully transfer the molecules to the rovibrational ground state ($v=0, J=0$), the first excited vibrational state ($v=1, J=0$), and the second excited rotational state ($v=0, J=2$) [Fig.~\ref{figure1}(b), (c) and (d), respectively]. Higher vibrational states, if necessary, could also be populated similarly.  

A challenge to produce \NaRb molecules in a single quantum state is the small hyperfine splittings. For the $X\,^1\Sigma^+$ state with zero total electronic angular momentum, the dominant contributions of the hyperfine structure (HFS) come from the atomic nuclear spins and their coupling with the nuclear rotation. With the nuclear spins of \Na and \Rb atoms, $I_{\rm Na}=I_{\rm Rb}=3/2$, there are $(2J+1)(2I_{\rm Rb}+1)(2I_{\rm Na}+1)=16$, 48 and 80 hyperfine levels for rotational states with $J=0$, 1 and 2, respectively. The overall frequency span of the HFS in each rotational state is only 2 to 6~MHz, while the intervals between adjacent hyperfine levels are even much smaller.

To prepare molecules in a single hyperfine level via STIRAP, two-photon linewidths narrower than the splittings between adjacent levels are needed. We achieve this with carefully chosen Rabi frequencies and long Raman pulses. The well-resolved HFS of the three rovibrational states in Fig.~\ref{figure1}(b), (c) and (d) are obtained with Raman laser pulses of 200~$\mu$s and maximum Rabi frequencies of $2\pi\times0.8$~MHz with corresponding two-photon linewidths of about 30~kHz. Under these conditions, nearly 90\% population transfer efficiencies can still be achieved. To eliminate the ac Stark shift induced by the trapping lights for a precise determination of the HFS line positions, the optical dipole trap is turned off during the two-photon Raman process.

Because of the angular momentum selection rules, not all the hyperfine levels can be reached by the two-photon Raman process. At $B=335.2$~G, energy levels of each vibrational state $v$ in the $X\,^1\Sigma^+$ potential can be represented in the uncoupled basis $\ket{J,m_J,m_I^{\rm Na},m_I^{\rm Rb}}$, with $m_J$, $m_I^{\rm Na}$ and $m_I^{\rm Rb}$ the projections of $\vec{J}$, $\vec{I}_{\rm Na}$ and $\vec{I}_{\rm Rb}$. During the transfer process, $M_F = m_J + m_I^{\rm Na} + m_I^{\rm Rb}$ is always a good quantum number. With $M_F=2$ for the Feshbach molecules and the polarization of $L_1$ fixed linearly along the $B$ field, $L_2$ with $\sigma^{\pm}$ and $\pi$ polarization can only access hyperfine levels with $M_F=1$, 2 and 3. For the $J=0$ rotational states with $M_F = m_I^{\rm Na} + m_I^{\rm Rb}$, there are six allowed hyperfine levels. In Fig.~\ref{figure1}(b) and (c), these hyperfine levels, including the absolute ground state in the ($v=0, J=0$) state labeled as $\ket{0,0,3/2,3/2}$ with $M_F=3$, are all resolved and thus can be populated with high quantum purity. Figure~\ref{figure1}(d) contains 22 of the total 29 accessible hyperfine levels of the $J=2$ state, which is much more complicated due to the nuclear spins and rotation coupling. It is also more challenging to populate some of the $J=2$ hyperfine levels with 100\% purity due to very close level spacings. As demonstrated in Fig.~\ref{figure1}(d), the situation can be improved with the help of the angular momentum selection rules by controlling the polarization of $L_2$.

\begin{figure}[bpt]
	\centering
	\includegraphics[width=0.45\textwidth]{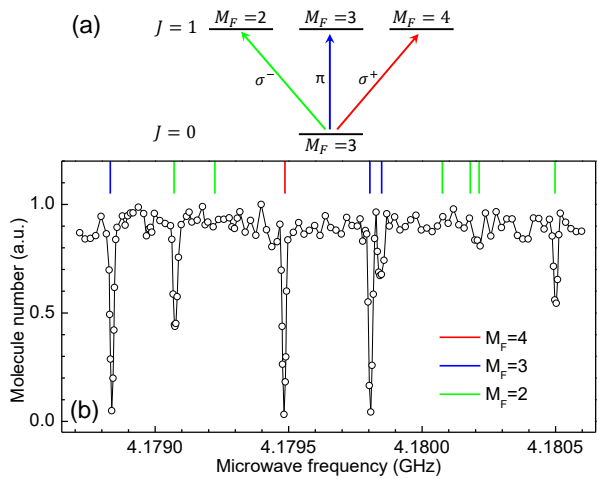}
	\caption{HFS of the $J=1$ rotational state probed by microwave spectroscopy. (a) depicts the allowed microwave transitions with different microwave polarizations starting from the $M_F = 3$ absolute ground state $\ket{0,0,3/2,3/2}$. (b) shows the observed HFS of the $J=1$ state manifested as loss of the absolute ground state population. The microwave pulse length is 50~$\mu$s which is shorter than a $\pi$ pulse for the power used. The colored vertical bars are the calculated positions of the relevant hyperfine levels of $J = 1$  using the fitted coupling constants in Table~\ref{table1}.}
	\label{figure2}
\end{figure}

Due to the parity selection rule, molecules cannot be prepared to the $J=1$ state directly via a STIRAP. This can be compensated by applying a microwave pulse driving the $J=0 \rightarrow J=1$ transition after transferring the molecules to $J=0$ state~\cite{ospelkaus2010controlling,will2016coherent,gregory2016controlling,park2017second}. This transition is typically very strong since the microwave couples directly to the large permanent electric dipole moment of ground-state \NaRb molecules~\cite{aymar2005calculation,guo2016creation}. To probe the HFS of the $J=1$ state, we detect the remaining $J = 0$ molecules with respect to the frequency after applying the microwave pulse. The frequency of the microwave is tuned around $2B_v\approx4.179$~GHz, with $B_v$ the rotational constant. To eliminate the ac Stark shift, the optical trap is turned off during the pulse. Typically, the microwave only changes the rotational level following $\Delta J=\pm1$ and $\Delta m_J=0,\pm1$, but will not flip the nuclear spins directly. However, because of the coupling between the nuclear spins and rotation, it becomes possible to manipulate the nuclear spins with electric dipole transitions. Figure~\ref{figure2}(b) shows the spectrum with 6 hyperfine levels of the $J=1$ state observed starting from the $M_F=3$ level of $J=0$.

%%%%%%%%%%%%%%%%%%%%%%%%%%

To describe the observed HFS, we use the Hamiltonian~\cite{brown2003rotational,aldegunde2008hyperfine,will2016coherent} 
\begin{equation}
	\centering
	H=H_{rot}+H_{hf}+H_Z,
	\label{equaiton1}
\end{equation}
with $H_{rot}=B_v J (J+1)$ the rotational splitting, $H_{hf}=\sum_{i}\bm{V}_i\cdot\bm{Q}_i+\sum_{i}c_i\bm{J}\cdot\bm{I}_i+c_3\bm{I}_{\rm Na}\cdot\bm{T}\cdot\bm{I}_{\rm Rb}+c_4\bm{I}_{\rm Na}\cdot\bm{I}_{\rm Rb}$ the hyperfine interactions, and $H_Z=-g_r\mu_N\bm{J}\cdot\bm{B}-\sum_{i}g_i(1-\sigma_i)\mu_N\bm{I}_i\cdot\bm{B}$ the Zeeman effects from the nuclear spins and rotation with $\mu_N$ the nuclear magneton, $g_r$ the rotational $g$-factor, $g_i$ the nuclear g-factors, and $\sigma_i$ the nuclear shielding tensor with $i={\rm Na, Rb}$. For $H_{hf}$, the first term is the electric quadrupole interactions associated with the coupling constants $(eqQ)_i$. This term dominates $H_{hf}$ for excited rotational states but vanishes for $J=0$. It is also the main cause of the nuclear spin and rotation mixing, which makes nuclear spin flip by microwave possible. The second term represents the direct rotation and nuclear spin coupling, which is typically rather small (Table~\ref{table1}). The last two terms describes the tensor and scalar interactions between the nuclear spins. At $B = 335.2$~G, for $J=0$ states, the nuclear Zeeman effect dominates all the other contributions, which results in a monotonous dependence of the hyperfine energy on $M_F$ [Fig.~\ref{figure1}(b) and (c)]. For $J > 0$, the electric quadrupole interactions are comparable to the nuclear Zeeman effects. Thus, the hyperfine energy also depends strongly on $m_J$ and the order of the HFS is more complicated. As summarized in Table~\ref{table1}, using this model to fit the observed hyperfine levels of the ($v=0;~J=0,1,2)$ states in Fig.~\ref{figure1}(b), (d) and Fig.~\ref{figure2}(b), the coupling constants for the hyperfine and Zeeman interactions are extracted. The fitting results are indicated by the color-coded vertical bars in Fig.~\ref{figure1}(b), (d) and Fig.~\ref{figure2}(b). The positions of the hyperfine levels from the measurement and the fitting agree with each other within 10~kHz. 

\begin{table}[bpt]
	\caption{Coupling constants in the molecular Hamiltonian [Eq.~(\ref{equaiton1})] for the $v=0$ state of \NaRb molecules. As $c_{\rm Na}$, $c_{\rm Rb}$ and $c_3$ are very small and barely affect the fitting results, they are fixed to the values provided in Ref.~\cite{aldegunde2017hyperfine} during the fitting. The theoretical value of $B_v$ is calculated with the experimental $X\,^1\Sigma^+$ potential in Ref.~\cite{pashov2005potentials}}
	\centering
	\begin{tabularx}{0.46\textwidth}{C C C}
			%		\centering
			\hline
			\hline
			Constant & Value & Reference\\
			\hline
			$B_v$ & 2.0896628(4)~GHz & This work\\
			& 2.08966~GHz & \cite{pashov2005potentials}\\
			$(eqQ)_{\rm Na}$ & -0.139(40)~MHz & This work\\
			& -0.132~MHz & \cite{aldegunde2017hyperfine}\\
			$(eqQ)_{\rm Rb}$ & -3.048(13)~MHz & This work\\
			& -2.984~MHz & \cite{aldegunde2017hyperfine}\\
			$c_{\rm Na}$ & 60.7~Hz & \cite{aldegunde2017hyperfine}\\
			$c_{\rm Rb}$ & 983.8~Hz & \cite{aldegunde2017hyperfine}\\
			$c_3$ & 259.3~Hz & \cite{aldegunde2017hyperfine}\\
			$c_4$ & 6.56(23)~kHz & This work\\
			& 5.73~kHz & \cite{aldegunde2017hyperfine}\\
			$g_{\rm Na}(1-\sigma_{\rm Na})$ & 1.484(1) & This work\\
			& 1.477 & \cite{arimondo1977experimental} \\
			$g_{\rm Rb}(1-\sigma_{\rm Rb})$ & 1.832(1) & This work\\
			& 1.827 & \cite{arimondo1977experimental}\\
			$g_r$ & 0.001(6) & This work\\
			\hline
			\hline
		\end{tabularx}
		\label{table1}
\end{table}

%%%%%%%%%%%%%%%%%%%%%%%%%%
\begin{figure}[bpt]
	\centering
	\includegraphics[width=0.45\textwidth]{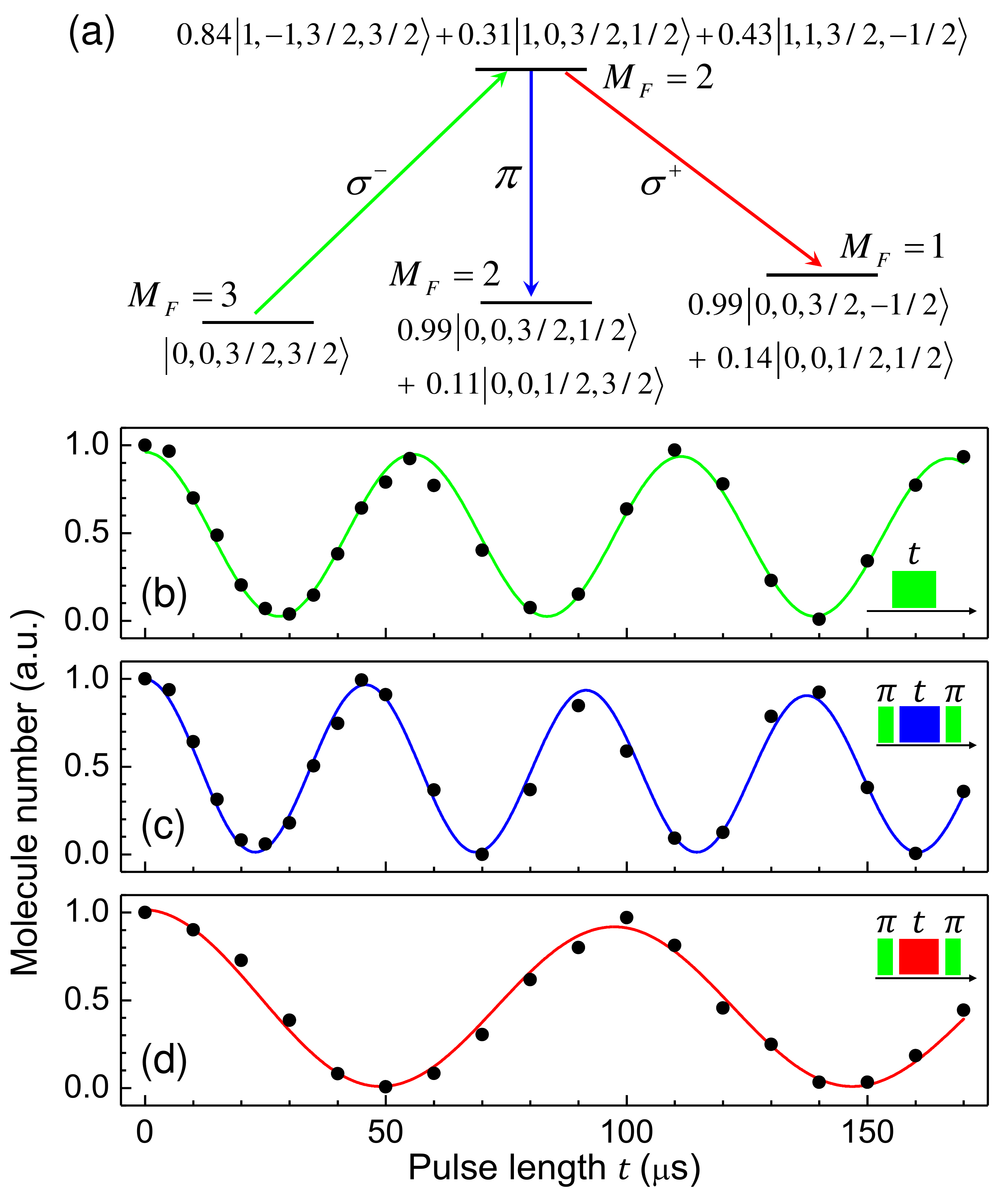}
	\caption{Coherent population transfer between the hyperfine levels of the $J=0$ and $J=1$ states with microwave pulses. (a) illustrates an example scheme for the coherent manipulation. The selected $M_F = 2$ hyperfine level of the $J=1$ state consists of a superposition of three nuclear spin components. (b), (c) and (d) show the observed Rabi oscillations for the three microwave transitions in (a). (c) and (d) are measured after transferring the population to $J=1$ from the $\ket{0,0,3/2,3/2}$ level with a $\pi$ pulse. }
	\label{figure3}
\end{figure}

From the fitting results, we also determine the mixing ratio of the $\ket{J,m_J,m_I^{\rm Na},m_I^{\rm Rb}}$ basis components in each hyperfine level. As shown by the example in Fig.~\ref{figure3}(a), the selected $M_F = 2$ hyperfine level in the ($v=0, J=1$) state has significant $\ket{1,-1,3/2,3/2}$, $\ket{1,0,3/2,1/2}$, and $\ket{1,1,3/2,-1/2}$ components. The $\ket{1,-1,3/2,3/2}$ component allows this level to be addressed from the absolute ground state $\ket{0,0,3/2,3/2}$. As shown by the coherent Rabi oscillations in Fig.~\ref{figure3}(b), with a microwave $\pi$ pulse, the population can be completely transferred from $J=0$ to $J=1$. 

In addition, the $\ket{1,0,3/2,1/2}$ and $\ket{1,1,3/2,-1/2}$ components of the ($v=0, J=1$) level in Fig.~\ref{figure3}(a) also enable us to manipulate the hyperfine levels of the $J=0$ state. As illustrated in Fig.~\ref{figure3}(a), after transferring molecules to the $J=1$ level with a $\pi$ pulse, a second microwave pulse can be applied to prepare molecules to $J=0$ hyperfine levels different from the initial one. The $M_F = 2$ and $M_F = 1$ hyperfine levels of the $J=0$ state also contain more than one nuclear spin components. The microwave transition strength is only between components of the same nuclear spin configurations of the two addressed levels, e.g., the $\ket{1,0,3/2,1/2}$ and the $\ket{0,0,3/2,1/2}$ components for the $\Delta M_F = 0$ transition. Figure~\ref{figure3}(c) and (d) show the Rabi oscillations when these transitions are driven resonantly. Another $\pi$ pulse to the $\ket{0,0,3/2,3/2}$ level is also applied subsequently to the second pulse in order to monitor the remaining population in the $J = 1$ level.

As expected, the transitions in Fig.~\ref{figure3} are all rather strong. With a moderate microwave power, appreciable Rabi frequencies of $2\pi\times18.1(1)$~kHz [Fig.~\ref{figure3}(b)], $2\pi\times21.8(1)$~kHz [Fig.~\ref{figure3}(c)] and $2\pi\times10.2(1)$~kHz [Fig.~\ref{figure3}(d)] can already be achieved. The ratios of the Rabi frequencies are in good agreement with the calculated mixing ratios of the different components for the levels involved in each corresponding transitions. These large Rabi frequencies will allow us to prepare molecules in different hyperfine levels as well as coherent superposition of two rotational states in very short time scales for various purposes in future investigations. 

In summary, we achieve full control over the internal degrees of freedom of \NaRb molecule, including vibrational, rotational and hyperfine levels, by combining a STIRAP and microwave spectroscopy. The hyperfine coupling constants of the vibrational ground state are determined with high accuracy by fitting the observed hyperfine levels of the three rotational states $J=0,~1,~2$. The control over the internal states of the molecules paves the way to investigate molecular collisions at different states, especially to compare molecular losses with or without the presence of chemical reactivity for $v=1$ and $v=0$ states~\cite{ye2017collisions}. Furthermore, as microwave transitions couple states with different parity, it can induce direct dipole-dipole interaction within the molecules, making it possible to engineer the molecular interactions and collision properties~\cite{buchler2007three,buchler2007strongly,gorshkov2008suppression}.

We thank Olivier Dulieu and Roman Vexiau for the valuable discussions.This work is supported by the COPOMOL project, jointly funded by Hong Kong RGC (grant NO. A-CUHK403/13) and France ANR (grant NO. ANR-13-IS04-0004-01). The Hong Kong team is also supported by the RGC General Research Fund (grant NO. CUHK14301815) and the National Basic Research Program of China (grant NO. 2014CB921403).

\bibliographystyle{apsrev4-1}
%\bibliography{reference}

%

\end{document}